\documentclass[10pt,conference,a4paper]{IEEEtran}
\makeatletter
 \def\ps@headings{%
 \def\@oddhead{\mbox{}\scriptsize\rightmark \hfil \thepage}%
 \def\@evenhead{\scriptsize\thepage \hfil \leftmark\mbox{}}%
 \def\@oddfoot{}%
 \def\@evenfoot{}}
 \makeatother
\ifCLASSINFOpdf
  \usepackage[pdftex]{graphicx}
  \graphicspath{{../pdf/}{../jpeg/}}
  \DeclareGraphicsExtensions{.pdf,.jpeg,.png}
\else
  \usepackage[dvips]{graphicx}
  \graphicspath{{../eps/}}
  \DeclareGraphicsExtensions{.eps}
\fi

\usepackage[cmex10]{amsmath}
\usepackage[tight,footnotesize]{subfigure}
\usepackage{url}

\usepackage{array}
\usepackage{tabularx}
\usepackage{amscd,amsbsy,amssymb,latexsym,url,bm}
\usepackage{epsfig,graphicx,subfigure}
\usepackage{wrapfig}
\usepackage{mathrsfs, euscript}
\usepackage[usenames,dvipsnames]{xcolor}
\usepackage{amsthm}
\usepackage{multicol}
\usepackage{lscape}
\usepackage{array}
\usepackage{tabularx}
\usepackage{pifont}
\usepackage{cite}
\usepackage{gensymb}
\usepackage{textcomp}
\usepackage{amsmath}
\usepackage[linesnumbered,vlined,boxed,ruled]{algorithm2e}
\usepackage{verbatim}
\usepackage{epstopdf}
\usepackage{listings}
\usepackage{fancyhdr}
\usepackage{enumitem}
\usepackage{longtable}
\usepackage{multirow}

\definecolor{background}{HTML}{EEEEEE}
\lstdefinelanguage{json}{
    basicstyle=\small\ttfamily,
    showstringspaces=false,
    breaklines=true,
    backgroundcolor=\color{background},
}

\usepackage{makecell}

\usepackage{listings}
\usepackage{xcolor}

\IEEEoverridecommandlockouts

\begin{document}
\SetAlFnt{\small}
\addtolength{\parskip}{0mm}
\setlength{\belowcaptionskip}{3pt}
\setlength{\abovecaptionskip}{3pt}
\setlength{\intextsep}{3pt}
\setlength{\floatsep}{3pt}
\setlength{\textfloatsep}{3pt}
\setlength{\dblfloatsep}{3pt}
\setlength{\dbltextfloatsep}{3pt}
\setlength{\belowdisplayskip}{3pt}
\setlength{\abovedisplayskip}{3pt}
\setlength{\columnsep}{0.241 in}


\title{Advancing IoT System Dependability: A Deep Dive into Management and Operation Plane Separation}

\author{\IEEEauthorblockN{Luoyao Hao, Shuo Zhang, Henning Schulzrinne}
\IEEEauthorblockA{Department of Computer Science, Columbia University, New York, NY, USA
}
{Email: \{lyhao, hgs\}@cs.columbia.edu}, sz3177@columbia.edu
}

%

\maketitle

\begin{abstract}
We propose to enhance the dependability of large-scale IoT systems by separating the management and operation plane. We innovate the management plane to enforce overarching policies, such as safety norms, operation standards, and energy restrictions, and integrate multi-faceted management entities, including regulatory agencies and manufacturers, while the current IoT operational workflow remains unchanged.  Central to the management plane is a meticulously designed, identity-independent policy framework that employs flexible descriptors rather than fixed identifiers, allowing for proactive deployment of overarching policies with adaptability to system changes. Our evaluation across three datasets indicates that the proposed framework can achieve near-optimal expressiveness and dependable policy enforcement. 
\end{abstract}

\section{Introduction}\label{section-intro}

Internet of Things (IoT) seeks to seamlessly connect everyday objects into a comprehensive system, enabling applications to run upon these interconnected devices to offer a wide array of services. As these systems evolve from individually configured devices and independently developed applications into intricate ecosystems, the inherent heterogeneity becomes increasingly apparent. Consider a smart building management system, which must integrate a range of devices and subsystems to automate these heterogeneous elements effectively, exemplified by a typical ventilation system dynamically adjusts based on environmental sensor data.

However, the integration and operation of individually developed components present significant challenges for large IoT systems, as any unexpected behavior could impact the broader ecosystem. With the increasing ease of achieving application-level integration and interaction, a plethora of conflicts and vulnerabilities emerge, underscoring the urgent need for improving system dependability~\cite{celik2019iotguard}. 
Dependability refers to the system's ability to provide reliably trustworthy services and encompasses critical attributes such as safety, availability, reliability, and maintainability. Among these attributes, safety is paramount~\cite{avizienis2004basic, eder2014towards}.

A critical question arises: how can we drastically enhance system dependability to ensure that devices operate as expected, or at least mitigate the impact of individual device failures, inappropriate actions, or inaccurate data? Defining expected behavior within such a system is challenging, necessitating the collaboration of various management entities, including regulatory agencies, manufacturers, and local management teams. For example, introducing a new heating pump should be regulated by the fire department and must comply with energy efficiency standards and operational guidelines. We argue that involving management entities and policies is crucial, not just advantageous. Rather than limiting flexibility, it actually elevates safe operation and integration, facilitating the effective functioning of large-scale IoT systems.

Traditionally, dependability is ensured through physical safeguards or spatial isolation before deployment~\cite{eder2014towards, fei2001safety}. In the context of IoT, 
recent studies mainly address the conflicts of human operations and automation rules~\cite{surbatovich2017some, ifttt, nacci2018buildingrules, celik2019iotguard, chaki2020conflict}. Additionally, a safety-oriented programming language is proposed to facilitate the development of safe applications~\cite {liang2015sift}. However, these efforts do not address the architectural shortcomings present in current IoT systems. While the need for a new framework has been acknowledged~\cite{xing2020reliability, eder2014towards}, it lacks comprehensive exploration.


In fact, resolving dependability issues in today's IoT operation is infeasible, since devices are allowed to autonomously make decisions that could adversely affect the broader system. Such architectural constraint, which hinders the implementation of management policy enforcement and the assessment of system-wide impacts, has to be addressed~\cite{zhang2024policy, hao2024poster}.

Hereby, we propose to separate the IoT management plane atop the operation plane, enabling more effective governance and nuanced policy enforcement in IoT systems. Achieving such a clean separation involves challenges, including detaching device identities and ensuring forward compatibility of policies. We design and develop the management plane comprising three essential components: a policy server, a policy engine, and a device directory. To allow for the proactive configuration of overarching policies, we design an Identity-Independent Policy (IIP) specification that focuses on flexible descriptors like relationships and properties, instead of fixed identities. This structure assures robust oversight, approving or overriding operations to align with policies for enhancing safety and compliance. We prototype the system, incorporating established standards and services, such as W3C TD~\cite{wot-thing-description}, Brick~\cite{brick-web}, and GOLDIE~\cite{hao2021infocom}, to avoid introducing fragmentation. Evaluations using real-world IFTTT and gas detection datasets show the system's significant expressiveness and effectiveness, achieving expression in 99\% of scenarios and detecting up to 94\% of hazards in unreliable settings.

We make the following contributions:
\begin{itemize}
\item We identify the necessity of decoupling management from operation in large IoT systems, to enhance dependability through the enforcement of overarching policies.
\item We design the management plane and the identity-independent policy specification, enabling management entities to configure wide-reaching policies proactively.
\item We prototype and open-source the management plane, demonstrating its expressiveness and effectiveness in improving system dependability\footnote{Open-sourced at \url{https://github.com/Halleloya/MP-IoT}}. 
\end{itemize}




\section{Motivation and Design Principles}\label{section-overview}
Current limitations and the evolvement of IoT systems motivate us to reconsider the architecture of IoT management.

\subsection{Motivation}
The motivation for separating management from operation lies on three pillars: (a) IoT system integration capabilities are stagnant due to increasing heterogeneity and potentially unsafe operations, (b) the distinction between operational rules and management policies is unclear, and (c) regulatory policies are not harnessed and management entities are not involved.  

IoT aims for system-wide automation and integration that simplifies tasks and streamlines operations. Yet, the growing mix of heterogeneous devices and independently developed systems underscore a clear need for structured governance to ensure that operations, data, and system changes adhere to overarching policies (e.g., safety norms, operation standards, energy restrictions, and expected behaviors), guarding against disruptions from irregular behaviors or data errors.


Unfortunately, today's IoT systems blur the lines between management policies, automation rules, and access controls, lacking a systematic approach to enforcing overarching policies that are often set by authoritative bodies like regulatory agencies, manufacturers, and development communities. Consider a smart building where opening a fire door follows policies from fire departments. Such policies must be configured without prior knowledge of specific device identities. However, IoT devices are designed to understand immediate actions or rule-based automation (e.g., IFTTT~\cite{ifttt}), and the distinction between user-created rules and overarching policies is unclear, which can introduce anomalies and potential hazards. Crucially, a distinct policy system, separate from operational controls, is needed to achieve ultimate dependability. This system should audit device operation conditions and guide outcomes, adhering to overarching policies for safety, compliance, adaptability, and resilience to system changes or evolution~\cite{surbatovich2017some}.

\subsection{Design Principles}
Our goal is to develop a systematic and future-proof strategy, adaptable to system changes and applicable across various IoT scenarios, incorporating broad overarching policies from management entities, to boost system dependability. We propose the following design principles for establishing a management plane for IoT. 
\subsubsection{Seamless User Experience} Users should experience smooth, uninterrupted use of IoT applications, without noticing the underlying complex management processes.
\subsubsection{Standard Compliance} System components should, where possible, adopt existing standards and ontologies.
\subsubsection{Flexible, Forward-Compatible Policies} Policies should be feasibly configured before device deployment, ensuring compatibility with future system modifications.
\subsubsection{Clean Separation Between Management and Operation} A clear division of responsibilities should be maintained, for independent innovation without cross-plane concerns. 
\subsubsection{Clean Separation Between Policies and Automation Rules} Policies and automation rules should be distinguished, each with its purpose, semantics, and implementation. 

In essence, developers and users should be able to concentrate on their perspective plane, with the operation plane offering consistent functionality and the management plane engaging in broad policies specified by various management entities. This also allows the management plane to function as an optional or incrementally integrated feature.

\section{Management and Operation Plane Separation}\label{section-separation}
In this section, we introduce the architectural separation between the management plane and the operation plane.


\subsection{System Architecture}

\begin{figure}[tbp]
    \centering
    \vspace{2mm}
    \includegraphics[width=0.49\textwidth]{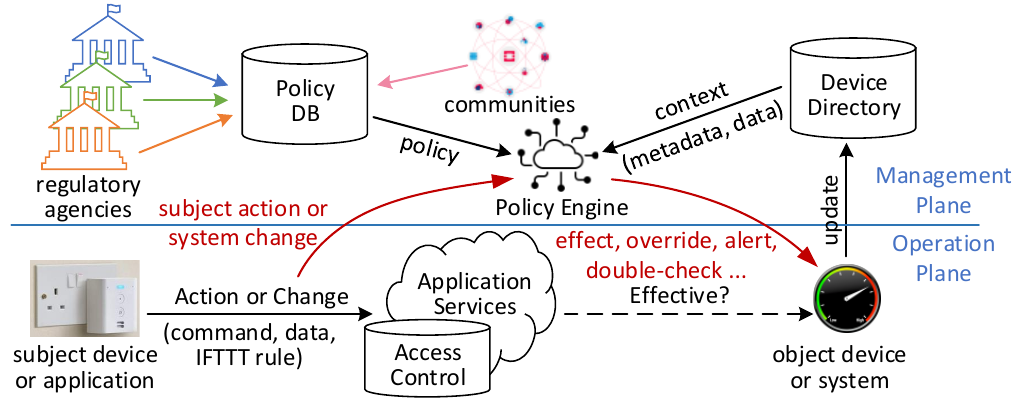}
    \vspace{-6mm}
    \caption{Separation between IoT management and operation.}
    \label{fig:separation}
\end{figure}

Aiming for a holistic and dependable IoT system through overarching policy enforcement, we propose an innovative architecture that features a distinct management plane on top of the current IoT operation, as shown in Fig.~\ref{fig:separation}.

The operation plane encapsulates the prevalent IoT system workflows where devices or applications can initiate commands, produce data, or propagate automation rules to the application service and IoT systems. Note that the access control still happens on the operation plane, which is not to be confused with the policy enforcement on the management plane, i.e., whether a device is authorized to initiate an action remains unchanged on the operation plane, while whether or how the action proceeds will be subject to the management policies. 
In contrast, the management plane operates as a governance logical layer, meticulously parsing and enforcing policies. It is composed of three system components: a Policy Database, a Policy Engine, and a Device Directory. The Policy Database serves as a repository for policies originating from various management bodies, such as regulatory agencies, manufacturers, and software communities. These policies might articulate adherence to fire codes, electrical guidelines, building codes, regional directives, and manufacture-specific operational rules in a structured expression format to be stored in the Policy Database.

Meanwhile, the Device Directory acts as a metadata repository, to provide device metadata and retrieves runtime data~\cite{hao2021infocom, hao2022dbac, aguzzi2024zion}. This data is subsequently requested by the Policy Engine, facilitating contextual policy evaluation. The Policy Engine, thus, as the core of the management plane, pulls policies and contexts (i.e., information requisitioned by the policies) to assess them against incoming actions or system alterations. The interplay between these elements results in a structured decision-making process, whereby the actual effect of an action or system modification is determined post-approval from the management plane. The management plane not only approves or denies what happens on the operation plane but also possibly overrides, enacts a double-check mechanism, or notifies related entities, ensuring that every action or system change is scrutinized against the needed policies, and thus affecting the system as enhanced safety and adherence to regulatory norms. This decoupling between management and operation plane fosters a meticulous examination of all actions against regulatory and safety policies, ensuring coherent, compliant, and safe IoT operations.

We reasonably assume that (1) devices keep their information updated in the directory, and (2) trained personnel of management entities are responsible for configuring policies. 

Traditional authorization policies are typically characterized by a ``allow/deny: subject–action–object'' syntax, which falls short of addressing the complex interactions and diverse decisions inherent in IoT ecosystems. For ease of expression, we structure the overarching policy model using three key terms, namely \textbf{change}, \textbf{context}, and \textbf{effect}, each playing a pivotal role in the policy framework.

\begin{description}
    \item[change:] the intended action or modification within the system, subject to policy evaluation.
    \item[context:] the set of conditions and information that inform the policy evaluation process.
    \item[effect:] the outcome of the policy evaluation, reflecting its impact on any devices or system components.
\end{description}

A \textbf{change} encompasses a broad spectrum of operations such as adjusting device configuration, introducing new data, scheduling events, or device replacement. The policy server scrutinizes the \textbf{change} to ensure it aligns with policies. A \textbf{context} encapsulates a variety of factors including device status, environmental variables, and temporal elements. These factors need to be obtained by the policy server to make precise decisions. An \textbf{effect} includes the enactment of decisions such as the approval or rejection of the changes, overriding commands, or triggering alerts. The \textbf{effect} is a tangible manifestation of the management plane, to enforce or alter the system's behavior under established policies.

\begin{figure}[tbp]
    \centering
    \includegraphics[width=0.42\textwidth]{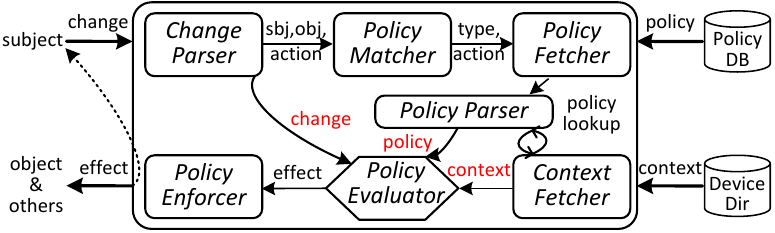}
    \vspace{-2mm}
    \caption{Policy engine functional modules.}
    \label{fig:policyserver}
\end{figure}

Within the dynamically shifting IoT environment, policies must adapt to anticipated and unforeseen changes, such as device substitutions or software updates. Such forward compatibility entails an identity-independent design, moving away from rigidly defined entity identifiers. Thus, we use device attributes and their interrelations, rather than fixed identifiers, to define policies, as specified in Section~\ref{section-specification}.


However, omitting device identifiers makes it challenging for the policy engine to find the exact set of policies to be evaluated, coupled with the complex IoT scenarios. Thus, a policy must provide: (1) whether the policy should be applied given a \textbf{change}, and (2) what \textbf{context} should be retrieved for evaluation. This is non-trivial as the device identities are not specified in the policy, diverging from typical authorization or management systems that explicitly link policies with specific devices and contexts with subjects or objects, and we need to avoid evaluating the entire set of policies for every \textbf{change}, ensuring efficiency and relevance in policy enforcement.

The functional modules of a policy engine, as depicted in Fig.~\ref{fig:policyserver}, begin their process when a \textbf{change} is initiated. The policy server needs to find a relevant set of policies, retrieve the \textbf{context} for each, evaluate them, and finally enforce the determined \textbf{effect} back to the operation plane. The \textbf{change} is formatted as ``subject–action–object'' (note that the \textbf{change} happens in the runtime, so it has identifiers, e.g., voice assistant X turns on light Y). 
The policy engine processes the \textbf{change}, extracting the incoming action and the device types of subjects and objects. The ``type-action'' pair guides the engine to filter through the policy database to retrieve a set of relevant policies, which can be a superset of the ideal policy set.
Subsequently, for each relevant policy, the engine fetches the \textbf{context}, probably of subject, object, and other affected entities, from the directory, as requested by the policy through flexible descriptors and relationships of devices. Compared to identity-based models, this approach, requiring policy lookup before evaluation, may be less intuitive but is necessary.



The separation of the management plane from IoT operations, with the identity-independent design, brings substantial benefits: (1) ensuring forward compatibility, (2) allowing for a compact and reusable policy set, and (3) facilitating the integration of various management entities into IoT ecosystems.

\section{Identity-Independent Policy Specification}\label{section-specification}
This section details the policy specification. Using flexible descriptors and relationships, we achieve compactness, portability, and forward compatibility. 


\subsection{Identity-Independent Policy}

We propose the Identity-Independent Policy (IIP). IIP identifies devices based on properties and inter-entity relationships, rather than specific identities, and thus achieves identity-independence. This is essential for enforcing overarching policies that are preferably configured before introducing devices (e.g., fire code compliance for any device at all times). With IIP, the management plane can achieve forward compatibility. 

Using IIP, a policy essentially defines the evaluation criteria for system changes (e.g., device actions, data inputs, event scheduling) and the necessary context for assessment. For example, ``\textless action: turn on, subject: \{type: AC\}, object: \{type: heater, status: on\}, relationship: AC.feed==heater.feed, result: warning\textgreater'' defines a policy to issue a warning when activating any AC in an area with a working heater, potentially violating energy-saving restrictions. 
Existing ontologies should preferably be adopted to define properties and relationships, such as W3C TD~\cite{wot-thing-description} and Brick~\cite{brick-web}. Thus, IIP does not reinvent ontologies, instead, it instantiates existing ontologies. 

\subsection{Policy Specification}

The specification defines elements for creating policies that are both human-understandable and machine-processable. Mandatory elements include id, action, relationship, and response, which are essential for policy formulation. Optional elements, such as type, affectedDevice, and expiration, offer additional features and support for diverse IoT scenarios. This structure is designed to facilitate straightforward and efficient policy management across different IoT use cases. Detailed definitions of these elements are provided as follows.

\begin{itemize}[leftmargin=*, align=left]
    \item id (String, Required): Unique identifier of the policy, formatted as ``[organization id]-[unique policy id assigned within the organization]''.
    \item description (String, Required): Human-readable description of the policy, formatted as ``[response]: description''.
    \item type (String, Optional): Category to which policy belongs. Used to specify different processing logic.
    \item subjectDevice (DeviceDetails, Optional): Device which initiates an action (defaults to the smartphone application).
    \item action (String, Required): The action that subjectDevice attempts to execute on objectDevice.
    \item objectDevice (DeviceDetails, Required): Device on which an action is executed.
    \item affectedDevice (Dict[string, DeviceDetails], Optional): Devices that may be affected by the action.
    \item relationship (Callable[Any], Required): Relationship between subjectDevice, objectDevice, and affectedDevice.
    \item assertion (Callable[Any], Optional): Conditions to be evaluated given the context.
    \item response (ResponseType(enum): [Approve, Deny, ...], Required): Response type when the assertion returns True.
    \item expiration (Datetime, Optional): Expiration date for the policy. Defaults to None (i.e., until deleted).
    \item alert (String, Optional): Alert message sent to the subjectDevice, objectDevice, or affectedDevice. 
    \item priority (Integer 0-9, Optional): Priority level of the Policy. 0 is the highest. Defaults to 6. 
\end{itemize}

The policy engine retrieves policies from the policy database, based on ``subjectDevice'', ``action'', and ``objectDevice''. It then combines ``affectedDevice'' and ``relationship'' to gather necessary context from the directory. The ``assertion'' ultimately checks conditions to determine the appropriate response to be enforced by the management plane.

The ``DeviceDetails'' class is defined below. The ``matchingAttribute'' aligns device attributes across various ontologies to simplified keywords used within the policy itself, seeking compatibility with both existing and future device ontologies.
\begin{itemize}[leftmargin=*, align=left]
    \item type (string, Required): Identifies the device type.
    \item matchingAttribute (Dict[string, string], Required): Maps device attributes to simplified keywords for policy processing.
\end{itemize}

\subsection{Simple Policy Example}
A simplified example is shown in Listing~\ref{iipexample} for illustration purposes (omitting some elements for space constraint). 

\begin{lstlisting}[language=json, caption={An example IIP for energy restriction: ``Double Check: turn on heater when AC is on for the same zone.'' It adopts both WoT and Brick ontologies to specify device attributes and relationships.}, captionpos=b, label=iipexample] 
/* omit id, description, type */
"action": "turn on",
"objectDevice": {
  "type": "heater",
  "matchingAttribute": {
    "type": "wot.type",
    "feeds": "brick.links.feeds"}},
"affectedDevice": {
  "ac": {
    "type": "ac",
    "matchingAttribute": {
      "type": "wot.type",
      "feeds": "brick.links.feeds",
      "on": "wot.property.on.status"}}},
"relationship": "set(objectDevice.feeds) & set(affectedDevice.ac.feeds)",
"assertion": "affectedDevice.ac.on == True",
"response": "Double Check",
"alert": "AC is on. Confirm to proceed."
\end{lstlisting} 

Incorporating functions into the ``relationship'' and ``assertion'' elements allows to express complex computational logic, e.g., for electrical safety, ``Deny: power a new device via a solar panel if the total power of currently active devices, plus the new device’s power, exceeds the solar panel’s capacity''.

\subsection{Guidelines for Creating New Policy}
To ensure efficient policy enforcement, we suggest these guidelines to adhere to for new policy creation:
\begin{enumerate}
    \item Use established ontologies for describing devices.
    \item Keep the assertion straightforward and concise. 
    \item Favor policies that incorporate confirmation or alert steps over direct approvals or rejections.
    \item Avoid assigning a short expiration duration. 
\end{enumerate}

\section{Prototype Implementation}
\label{section-implementation}

This section outlines our prototype system's implementation, discussing the options considered and the decisions made for the management plane.


\subsection{Management Plane Implementation}
Adopting an identity-independent approach for policy specification requires accommodating a wide variety of ``things'' and ensuring forward compatibility by focusing on flexible attributes and relationships over fixed identities. Besides, context retrieval requires the pre-inspection of policies prior to evaluation. Traditional authorization languages, with their binary, permission-based approach and pre-defined context requirements, are often ill-suited to encompassing the richness and complexity of IoT. 


Thus, we believe a more generalized policy language, closer to conventional programming languages, is required to encapsulate the intricacies and variances.  We use Python to develop the IIP specification and the policy system that are both expressive and flexible. The system requires type annotations for variable restrictions, and a validation tool is being developed. Essentially, policy creation mirrors object instantiation of the IIP specification class implemented in Python, familiarizing developers and administrators with the platform. This process is guided by a template, including examples of different policy logic, and bounded by class definitions, which streamlines the policy creation. A detailed comparison of common policy-specific language, including Rego~\cite{rego}, Casbin~\cite{casbin}, Sentinel~\cite{sentinel}, and Polar~\cite{polar}, with our implementation of IIP is shown in Table~\ref{table:lanCompare}.

\begin{table}[tbp]
\centering
\caption{Comparison of policy languages.}
\vspace{-2mm}
\centering
\renewcommand{\arraystretch}{1.05}
\begin{tabular}{ |l|l|l|l|l|l| } 
 \hline
\textbf{} & Rego & Casbin & Sentinel & Polar & IIP \\ \hline
Expressiveness & $\bullet$ $\bullet$ & $\bullet$ & $\bullet$ $\bullet$    & $\bullet$ & $\bullet$ $\bullet$  $\bullet$     \\
Brevity  & $\bullet$ $\bullet$ $\bullet$ & $\bullet$ & $\bullet$ & $\bullet$ $\bullet$ & $\bullet$ $\bullet$ \\
Flexibility & $\bullet$ $\bullet$  & $\bullet$ $\bullet$  & $\bullet$ & $\bullet$ $\bullet$ & $\bullet$ $\bullet$ $\bullet$    \\
Portability & $\bullet$ $\bullet$ $\bullet$    & $\bullet$    & $\bullet$ & $\bullet$ $\bullet$      & $\bullet$ $\bullet$ $\bullet$    \\
Pre-Inspection & $\bullet$ & -     & -      & -      & $\bullet$ $\bullet$ $\bullet$    \\ \hline
\end{tabular}
\label{table:lanCompare}
\end{table}

Alongside the policy engine, we implement Restful APIs using Flask framework for policy operations and requests handling. We utilize MongoDB for both directory and policy databases. For the device directory, we implement a centralized, single local directory for our prototype, based on~\cite{hao2021infocom}, to facilitate device detail provision, which adopts the W3C Thing Description, a JSON-based ontology for detailing properties, actions, and events of IoT devices~\cite{wot-thing-description}.


\subsection{Communication Between Planes}
\begin{figure}[htbp]
    \centering
    \vspace{1mm}
    \includegraphics[width=0.42\textwidth]{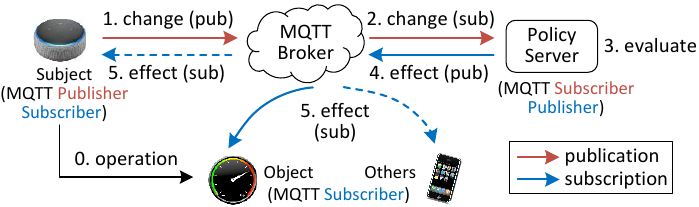}
    \vspace{-2mm}
    \caption{MQTT-based implementation for communication between planes.}
    \label{fig:mqttworkflow}
\end{figure}

While communication between the management plane and the operation plane mainly relies on the protocol support of devices, we practice it using MQTT for our prototype. As illustrated in Fig.~\ref{fig:mqttworkflow}, subject devices publish changes that the policy server subscribes to, and the policy server publishes effects for subscription by all relevant devices, possibly including the subject devices or other devices. This publish/subscribe model facilitates efficient message exchange across the system, ensuring that all necessary devices are kept informed of relevant changes and effects.

\begin{figure*}[tbp]
\begin{minipage}[b]{0.3\linewidth}
\centering
    \includegraphics[width=1\textwidth]{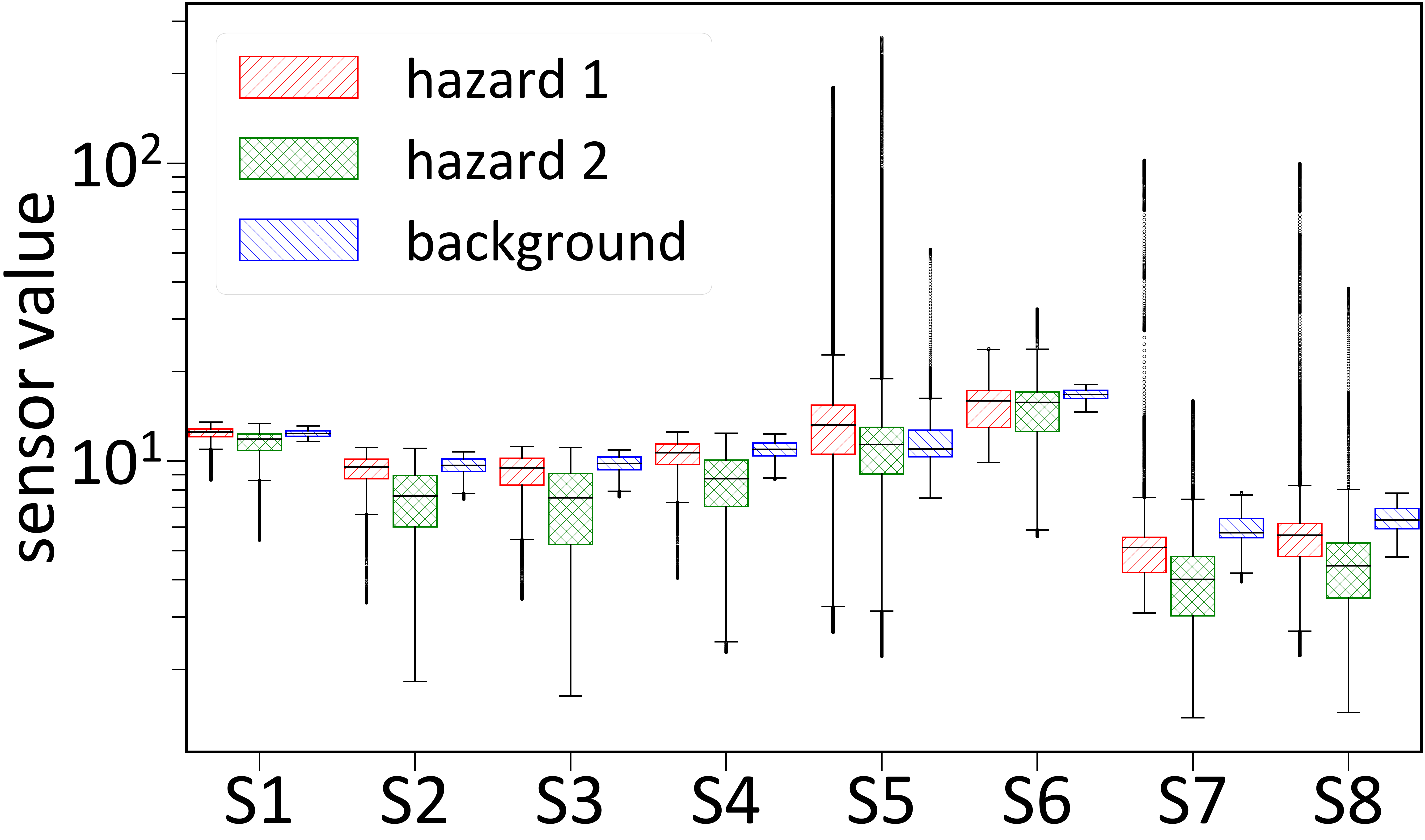}\\
    \caption{Sensor data distribution.}
    \label{fig-datadistribution}
\end{minipage}
\hspace{7mm}
\begin{minipage}[b]{0.3\linewidth}
\centering
    \includegraphics[width=1\textwidth]{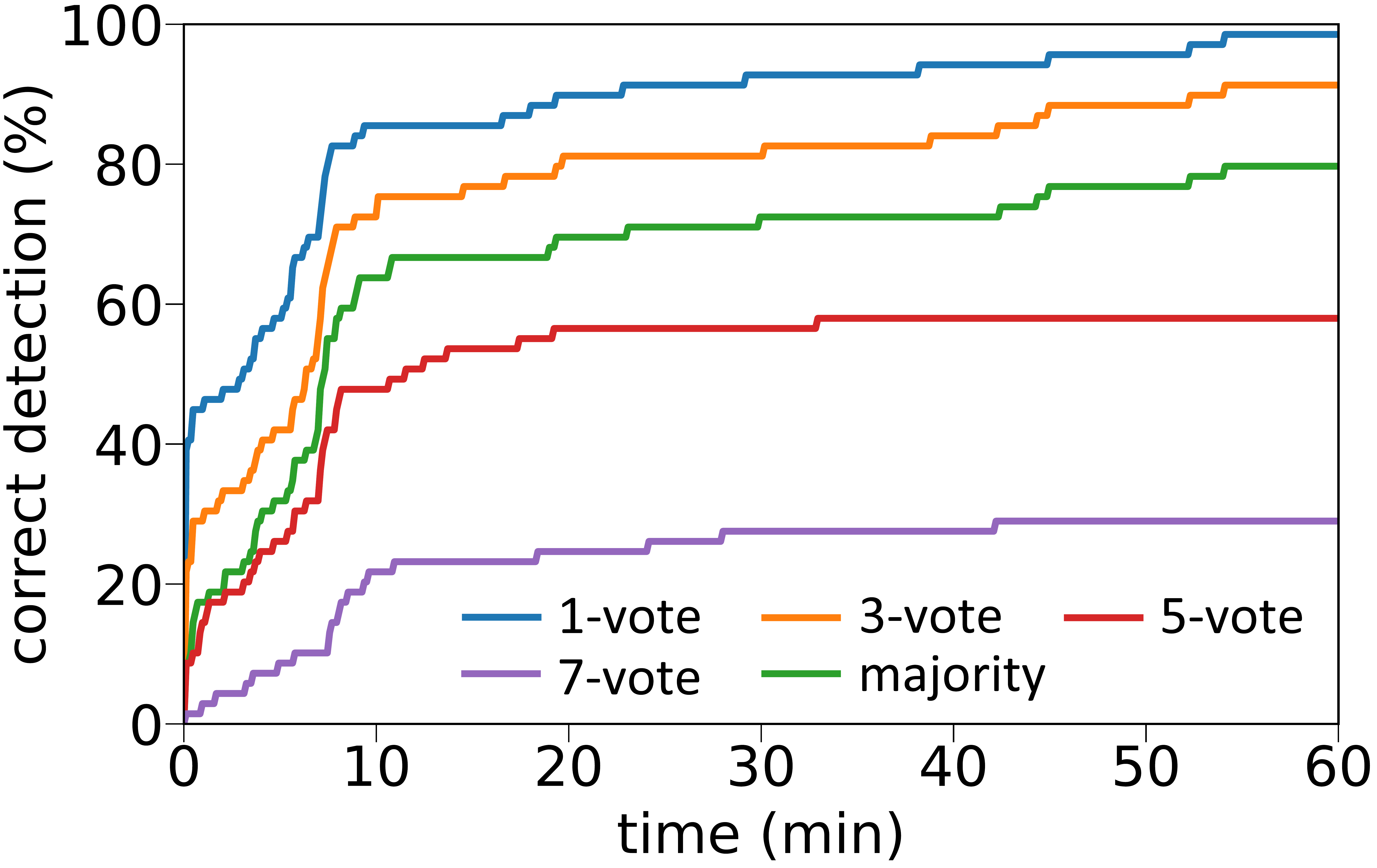}\\
    \vspace{-2mm}
    \caption{Correct detection as time elapses.}
    \label{fig-detection}
\end{minipage}
\hspace{6mm}
\begin{minipage}[b]{0.3\linewidth}
\centering
    \includegraphics[width=1\textwidth]{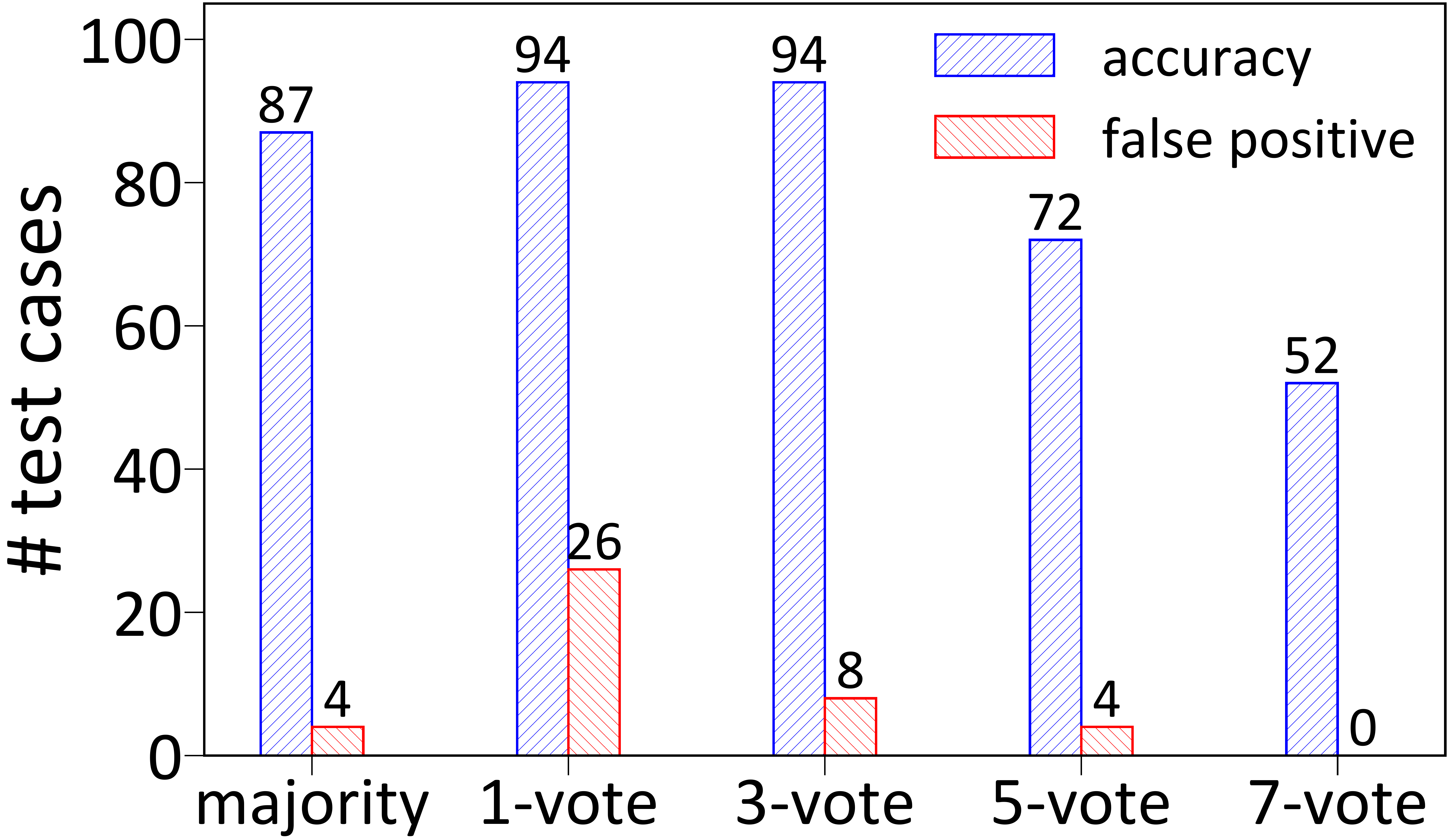}\\
    \caption{Accuracy and false positive.}
    \label{fig-accuracy}
\end{minipage}
\end{figure*}

\section{Evaluation}
\label{section-evaluation}
In this section, we evaluate the expressiveness of the policy specification and the advantages of the management plane.
 

\subsection{IIP Expressiveness}
We evaluate the expressiveness of the IIP specification using two real-world IoT IFTTT datasets~\cite{liu2019remediot, yu2021analysis}, with 201 and 2788 IFTTT recipes, respectively. Due to the lack of structured IoT policy datasets, these IFTTT collections, capturing complex IoT interactions, actually serve as the best available sources for assessing IIP comprehensiveness, although, in principle, IFTTT focuses on the operation plane. 

Table~\ref{table-expressive} shows the expressiveness capability of IIP when applied to the IFTTT datasets. It highlights particularly great expressiveness for policies pertaining to safety and energy. We also notice a substantial proportion requires the integration of external APIs for functionalities like time or location. The data indicates that the IIP can effectively represent approximately 99\% of the IFTTT recipes within the datasets, signifying its applicability in diverse IoT scenarios.

\begin{table}[htbp]
\centering
\caption{Evaluation of expressiveness using datasets}
\vspace{-2mm}
\label{table-expressive}
\renewcommand{\arraystretch}{1.05}
\resizebox{0.95\linewidth}{!}{
\begin{tabular}{|l|l|c|c|c|}
\hline
Dataset & Policy Type & Expressible & Require API & Inexpressible \\ \hline
\multirow{3}{*}{\cite{liu2019remediot}} & Safety/Energy & 62.50\% & 37.50\% & 0.00\% \\ 
 & Others & 56.52\% & 41.61\% & 1.86\% \\ 
 & Total & 57.71\% & 40.80\% & 1.49\% \\ \hline
\multirow{3}{*}{\cite{yu2021analysis}} & Safety/Energy & 64.01\% & 35.56\% & 0.43\% \\ 
 & Others & 51.75\% & 47.58\% & 0.67\% \\ 
 & Total & 54.84\% & 44.55\% & 0.61\% \\ \hline
\end{tabular}
}
\end{table}

\subsection{Policy Server Advantage}

The strength of the management plane lies in its ability to implement nuanced policies that offer adaptability beyond what is typically embedded by manufacturers. To demonstrate its effectiveness, we employ a dataset~\cite{huerta2016online} that originally captures time series from eight gas sensors to detect two stimuli: wine and banana. We interpret wine and banana as two hazards to be detected, simulating a highly unreliable scenario. 
Fig.~\ref{fig-datadistribution} presents the data distribution for sensors $S1$ to $S8$. The overlapping and the variability suggest an unreliable setting where hazards are not easily detected or separated from the background data. We then model each sensor's data into Gaussian distributions $\sim \mathcal{N}(\mu, \sigma^2)$ for each condition (i.e., hazard 1, hazard 2, and background). Given a data point, a sensor determines the current condition based on the maximum likelihood. Note that precise modeling and prediction are neither necessary nor within our scope. 

We consider policies based on consensus among sensors using voting mechanisms (i.e., ``Alert: if X number of sensors detect a hazard''). Figure~\ref{fig-detection} displays the evolution of hazard detection rates over time, revealing that policies with more lenient criteria, such as the 1-vote and 3-vote policies, achieve higher success rates of 98.55\% and 91.30\%, respectively, compared to 81.16\% achieved by majority voting. These policies also expedite detection, being 1.87 and 1.38 times faster, accordingly. Fig.~\ref{fig-accuracy} shows the accuracy and false positive results, including identifying both hazards and background, across different voting policies. Compared to majority voting, the 3-vote policy identifies 7 more cases accurately with a marginal elevation of 4 more false positives.

For system dependability, ensuring accurate and prompt issue detection is often prioritized over avoiding false positives. The evaluation demonstrates that, as the management plane evolves towards more adaptable policy enforcement, it facilitates nuanced policy applications in integrated systems.

\section{Limitation and Discussion}
\label{section-discussion}
\vspace{-1mm}
\paragraph*{Vulnerability} We implement the policy specification using a generic programming language, due to limited support in domain-specific languages beyond authorization scenarios. 
However, adopting advanced domain-specific languages with more capabilities remains viable, e.g., Pkl~\cite{pkl}.

\paragraph*{Policy Creation} Converting real-world policies (e.g., fire code) into machine-readable formats presents significant challenges. We discuss the limitations and solutions based on Large Language Models in~\cite{zhang2024policy}.



\section{Conclusion}\label{section-Conclusion}
\vspace{-1mm}
To enhance IoT system dependability, we address the architectural deficiency in incorporating management entities and policies, by innovatively separating the management and operation plane. The management plane, comprising a policy engine, a policy database, and a device directory, features elaborate identity-independent policy processing, ensuring forward compatibility, adaptability, and portability. We prototype using established standards and tools. Through evaluations with datasets, we demonstrate significant advantages in policy expression and system management, setting the stage for dependable, large-scale IoT ecosystems.

\section*{Acknowledgement}
\vspace{-1mm}
This work is supported by the National Science Foundation under grants CNS-1932418 and EEC-2133516. The authors would like to thank Jan Magnusson and Florian Nußbaum for their help in understanding policy languages. 

\bibliographystyle{IEEEtran}
\bibliography{bibs/general,bibs/safety}

@inproceedings{hao2021infocom,
  title={GOLDIE: Harmonization and Orchestration Towards a Global Directory for {IoT}},
  author={Hao, Luoyao and Schulzrinne, Henning},
  booktitle={IEEE INFOCOM},
  year={2021}
}

@inproceedings{hao2022dbac,
  title={{DBAC}: Directory-based access control for geographically distributed {IoT} systems},
  author={Hao, Luoyao and Naik, Vibhas and Schulzrinne, Henning},
  booktitle={IEEE INFOCOM},
  pages={360--369},
  year={2022}
}

@inproceedings{zhang2024policy,
  title={Policy Enforcement for {IoT}: Complexities and Emerging Solutions},
  author={Zhang, Shuo and Hao, Luoyao and Schulzrinne, Henning},
  booktitle={IEEE IPCCC},
  year={2024},
}

@inproceedings{hao2024poster,
  title={Poster: Identity-independent {IoT} for overarching policy enforcement},
  author={Hao, Luoyao and Schulzrinne, Henning},
  booktitle={IEEE S\&P Workshops},
  year={2024}
}

@inproceedings{aguzzi2024zion,
  title={{ZION}: A Scalable {W3C Web of Things} Directory},
  author={Aguzzi, Cristiano and Gigli, Lorenzo and Zyrianoff, Ivan and Roffia, Luca},
  booktitle={IEEE CCNC},
  year={2024}
}

@misc{ifttt,
    title={{IFTTT}: Automation for Business and Home},
    key={ifttt},
    url={https://ifttt.com/},
    note={Accessed: 2023-09-28}
}

@misc{brick-web,
  title={Brick Schema},
  key = {brickschema},
  year={2016},
  url={https://brickschema.org/},
  note={Accessed: 2023-05-02}
}

@misc{wot-thing-description,
  title={{Web of Things} Thing Description},
  key = {wottd},
  year={2023},
  url={https://www.w3.org/TR/wot-thing-description11/},
  note={Accessed: 2024-01-02}
}

@misc{rego,
    title={{OPA}: Open Policy Agent},
    key={opa},
    url={https://www.openpolicyagent.org/},
    note={Accessed: 2023-09-28}
}

@misc{casbin,
    title={Casbin: An authorization library},
    key={casbin},
    url={https://casbin.org/},
    note={Accessed: 2023-09-28}
}

@misc{sentinel,
    title={Sentinel by {HashiCorp}},
    key={sentinel},
    url={https://www.hashicorp.com/sentinel},
    note={Accessed: 2023-09-28}
}

@misc{polar,
    title={{OSO}: Authorization as Code},
    key={oso},
    url={https://www.osohq.com/},
    note={Accessed: 2023-09-28}
}

@misc{pkl,
    title={Configuration that is Programmable, Scalable, and Safe},
    key={pkl},
    url={https://pkl-lang.org/},
    note={Accessed: 2024-03-17}
}

@article{huerta2016online,
  title={Online decorrelation of humidity and temperature in chemical sensors for continuous monitoring},
  author={Huerta, Ramon and Mosqueiro, Thiago and Fonollosa, Jordi and Rulkov, Nikolai F and Rodriguez-Lujan, Irene},
  journal={Elsevier CILS},
  volume={157},
  pages={169--176},
  year={2016}
}

@inproceedings{celik2019iotguard,
  title={{IoTGuard}: Dynamic Enforcement of Security and Safety Policy in Commodity {IoT}},
  author={Celik, Z Berkay and Tan, Gang and McDaniel, Patrick D},
  booktitle={NDSS},
  year={2019}
}

@article{avizienis2004basic,
  title={Basic concepts and taxonomy of dependable and secure computing},
  author={Avizienis, Algirdas and Laprie, J-C and Randell, Brian and Landwehr, Carl},
  journal={IEEE TDSC},
  volume={1},
  number={1},
  pages={11--33},
  year={2004}
}

@inproceedings{eder2014towards,
  title={Towards the safety of human-in-the-loop robotics: Challenges and opportunities for safety assurance of robotic co-workers'},
  author={Eder, Kerstin and Harper, Chris and Leonards, Ute},
  booktitle={IEEE RO-MAN},
  pages={660--665},
  year={2014}
}

@inproceedings{chaki2020conflict,
  title={A conflict detection framework for {IoT} services in multi-resident smart homes},
  author={Chaki, Dipankar and Bouguettaya, Athman and Mistry, Sajib},
  booktitle={IEEE ICWS},
  pages={224--231},
  year={2020}
}

@article{xing2020reliability,
  title={Reliability in {Internet of Things}: Current status and future perspectives},
  author={Xing, Liudong},
  journal={IEEE IoT-J},
  volume={7},
  number={8},
  pages={6704--6721},
  year={2020}
}

@article{fei2001safety,
  title={The safety issues of medical robotics},
  author={Fei, Baowei and Ng, Wan Sing and Chauhan, Sunita and Kwoh, Chee Keong},
  journal={Elsevier RESS},
  volume={73},
  number={2},
  pages={183--192},
  year={2001}
}

@inproceedings{liang2015sift,
  title={{SIFT}: building an {Internet} of safe things},
  author={Liang, Chieh-Jan Mike and Karlsson, B{\"o}rje F and Lane, Nicholas D and Zhao, Feng and Zhang, Junbei and Pan, Zheyi and Li, Zhao and Yu, Yong},
  booktitle={IEEE/ACM IPSN},
  pages={298--309},
  year={2015}
}

@article{nacci2018buildingrules,
  title={{BuildingRules}: A Trigger-Action--Based System to Manage Complex Commercial Buildings},
  author={Nacci, Alessandro A and Rana, Vincenzo and Balaji, Bharathan and Spoletini, Paola and Gupta, Rajesh and Sciuto, Donatella and Agarwal, Yuvraj},
  journal={ACM TCPS},
  volume={2},
  number={2},
  year={2018}
}

@inproceedings{surbatovich2017some,
  title={Some recipes can do more than spoil your appetite: Analyzing the security and privacy risks of {IFTTT} recipes},
  author={Surbatovich, Milijana and Aljuraidan, Jassim and Bauer, Lujo and Das, Anupam and Jia, Limin},
  booktitle={ACM WWW},
  pages={1501--1510},
  year={2017}
}

@inproceedings{liu2019remediot,
  title={{RemedIoT}: Remedial actions for {Internet-of-Things} conflicts},
  author={Liu, Renju and Wang, Ziqi and Garcia, Luis and Srivastava, Mani},
  booktitle={ACM BuildSys},
  pages={101--110},
  year={2019}
}

@inproceedings{yu2021analysis,
  title={Analysis of {IFTTT} recipes to study how humans use {Internet-of-Things (IoT)} devices},
  author={Yu, Haoxiang and Hua, Jie and Julien, Christine},
  booktitle={ACM BuildSys},
  pages={537--541},
  year={2021}
}
\end{document}